\documentstyle[aps,prl,epsf,epsfig,twocolumn,float]{revtex}

\begin{document}
\draft

\wideabs{
\title{Anisotropy of superconducting MgB$_{2}$ as seen in electron spin
resonance and magnetization data
}
\author{F. Simon, A. J\'{a}nossy, T. Feh\'{e}r, F. Mur\'{a}nyi}
\address{Budapest University of Technology and Economics, Institute
of Physics\\
and Solids in Magnetic Fields Research Group of the Hungarian Academy of\\
Sciences, H-1521 Budapest, PO BOX 91, Hungary}
\author{S. Garaj, L. Forr\'{o}}
\address{Laboratoire de Physique des Solides Semicristallins, IGA
Department de\\
Physique, Ecole Polytechnique Federal de Lausanne, 1015 Lausanne, Switzerland}
\author{C. Petrovic, S.L. Bud'ko, G. Lapertot\cite{lapertot},  V.G. Kogan, P.C.
Canfield}
\address{Ames Laboratory, U.S. Department of Energy and Department of Physics\\
and Astronomy, Iowa State University, Ames, Iowa 50011}
\date{\today}
\maketitle

\begin{abstract}
We have observed the conduction electron spin resonance (CESR) in  fine
powders of MgB$_{2}$ both in the superconducting and normal states. The Pauli susceptibility
is $\chi _{s}$=2.0$\cdot $10$^{-5}$ emu/mole in the temperature range of 450 to 600 K.
The spin relaxation rate has an anomalous temperature dependence.
The CESR measured below $T_{c}$ at
several frequencies suggests that MgB$_{2}$ is a strongly anisotropic
superconductor with the upper critical field, $H_{c2}$, ranging between 2
and 16 T.  The high-field reversible magnetization data of a randomly oriented
powder sample are well described assuming that MgB$_{2}$ is an anisotropic
superconductor with
  $H_{c2}^{ab} / H_{c2}^{c} \approx $ 6--9.
\end{abstract}


}
\narrowtext


Following the recent discovery of superconductivity in MgB$_{2}$\cite{akimitsu} 
several of its fundamental properties have been established. MgB$_{2}$ 
is a type II superconductor with $\lambda \approx 140\,$nm \cite
{finnemorePRL} and the upper critical field $H_{c2}\approx 16\,$T\cite{wirecondmat}. 
The question to what degree this superconductor is
anisotropic is still unresolved, the reason being the lack of single
crystals of size sufficient for direct measurements. The anisotropy is an
important characteristics both for the basic understanding of this material
and for applications; enough to mention that the anisotropy strongly affects 
the pinning and critical currents.

An anisotropic or multi-component superconducting gap was inferred from a
number of indirect measurements and was suggested in several theoretical
descriptions of MgB$_{2}$ \cite
{junod,hinksSH,shulga,bascones,liu,haas,chen,pronin}. For partially oriented
crystallites, the anisotropy ratio is reported as $\gamma
=H_{c2}^{ab}/H_{c2}^{c}=1.73$ \cite{brasil}, for {\it c}-axis oriented films
$\gamma \approx $2 \cite{patnaik} was found. In this Letter, we report
estimates of the anisotropy parameter $\gamma $ as high as 6--9, based on two
independent techniques which utilize properties of random powders.

Conduction Electron Spin Resonance (CESR) is commonly used to determine the
spin susceptibility, $\chi _{s}$, and the spin relaxation rate, $T_{1}^{-1}$, 
in normal metals. The mechanisms inducing conduction electron spin-lattice
relaxation are similar to those of momentum relaxation: they are both
related to phonons at high temperatures and to the impurity scattering at
low-$T$. In the mixed state of superconductors, CESR is observable due to
the normal electron states localized in vortex cores and due to
quasiparticle excitations over the gap (at finite temperatures).
Surprisingly, in powders of MgB$_{2}$ we also observe the normal phase CESR
signal at low $T$'s and in fields well below the reported upper critical
field of 16 T\cite{wirecondmat}. The data suggest that $H_{c2}$ of MgB$_{2}$ may be strongly
anisotropic.

We have used isotopically pure Mg$^{11}$B$_{2}$ ( $T_{c}$=39.2 K) samples
from the same batch as reported elsewhere \cite{budkoPRL}. The original
sample consisted of 100 $\mu$m large aggregates of small grains. The samples were
thoroughly ground in a mortar to crush the aggregates. Most of the resulting
grains were between 0.5 and 5 $\mu$m in size and were separated by mixing
into ESR silent high vacuum grease or SnO$_{2}$. Crushing the aggregates
increased the CESR signal intensity limited by small microwave penetration
but did not affect the superconducting properties of the samples: dc
magnetization measurements confirmed that $T_{c}(H)$,  the transition width
and shielding fraction remained unchanged. 

ESR experiments were performed at 9, 35, 75, 150 and 225 GHz at the
corresponding resonance magnetic fields of 0.33, 1.28, 2.7, 5.4 and 8.1 T.
The spin susceptibility was measured by calibrating the 9 GHz spectrometer
(Bruker ESP 300) against CuSO$_{4}\cdot $5H$_{2}$O. The 9 GHz spectrometer
uses a microwave resonant cavity and the so-called vortex noise generated by
the magnetic field modulation prohibits ESR measurements in the superconducting
state below the irreversibility line \cite{andrasphysC}. The High Field ESR
spectrometer (Budapest HF-ESR lab, 35 GHz and higher frequencies) does not
utilize a resonant cavity thus avoids vortex-noise. The $g$-factor was
measured with respect to diphenyl-picryl-hydrazyl ($g$=2.0036) and Mn/MgO
($g$=2.0009).


The CESR at 9 GHz and above 500 K has the antisymmetric Lorentzian
absorption derivative lineshape characteristic of a relaxationally
broadened ESR and homogeneous excitation (Fig. \ref{linewidth} inset). We
find $g$=2.0019$\pm $0.0001 for the $g$-factor at 40 K at both 35 GHz and 9
GHz. At 300 K and 9 GHz we get $g$=2.001$\pm $0.001. The CESR intensity is
temperature independent between 450 and 600 K and the paramagnetic spin
susceptibility is $\chi _{s}$=(2.0$\pm $0.3)$\cdot $10$^{-5}$ emu/mole.
Assuming negligible electron-electron correlations, the density of states
(DOS) at the Fermi level is 0.6 states/eV in agreement with band structure
calculations\cite{mazin}. At lower temperatures, the size of larger grains
(s = 5 $\mu$m) becomes comparable or larger than the skin depth, $\delta $
= ($\rho $/$\pi \mu _{0}f$)$^{1/2}$, and the observed CESR intensity
gradually decreases. Here $\rho $ denotes the specific resistivity of MgB$_{2}$, 
$\mu _{0}$ is the vacuum permeability, and $f$ is the ESR frequency.
At 40 K, the ESR intensity is $\sim $25\% of the high temperature value in
agreement with the decrease in $\delta $ estimated from the resistance (Fig. 
\ref{linewidth}). At 300 K and 40 K the calculated skin depths at $f$= 9 GHz
are $\delta $ = 1.6 $\mu$m and 0.3 $\mu$m, respectively, using values of $\rho $ 
measured on dense MgB$_{2}$ wires\cite{canfieldPRL}.

Below 400 K, deviations from the antisymmetric lineshape appear. In most
cases, a $T$ dependent mixture of Lorentzian derivative absorption and
dispersion components simulates well the observed line. Figure \ref
{linewidth} shows the $T$ dependence of the CESR linewidth in the normal
state. The Lorentzian lineshape is a signature of a homogeneous line
broadening; in this case the half width at half maximum of the Lorentzian absorption
line is $w=1/\gamma _{e}T_{1}$, where $T_{1}$ is the spin-lattice
relaxation time, and $\gamma _{e}$ is the electronic gyromagnetic factor. 
The electron mean free path, $\ell$, is about 0.06 $\mu$m
at 40 K\cite{canfieldPRL}, thus $\ell \ll\delta$ and the normal skin effect
determines the excitation. At 40 K, the spin mean free path\cite{winter}, 
$\delta_{eff} = 1/3v_{F}(T_1\tau )^{1/2}\approx 4\,$ $\mu$m is comparable to
the maximum grain size, 5 $\mu$m, and the conduction electron magnetization
is homogeneous. There is a field dependent residual linewidth at low $T$,
followed by a strong increase with temperature and a broad maximum at 450 K.
At 40 K and 9 GHz the linewidth is narrow, 15 G. At 40 K\ and 225 GHz the
line is inhomogeneous; it is broadened to about 35 G. The $T$ dependent
contribution to linewidth is independent of the magnetic field.

As expected for a light metal \cite{yafet}, MgB$_{2}$ has a $g$-factor of
2.0019 close to the free electron value of 2.0023 and a temperature
dependent linewidth (proportional to 1/$T_{1}$) which follows the
resistance below 200 K. However, the maximum in 1/$T_{1}$ observed in MgB$_{2}$ at 450
K has no analogue among pure metals.

Below $T_{c}$ the CESR changes dramatically, see Fig. \ref{spectra}. At 35
GHz (1.28 T) there is a single line at all temperatures, and a large $T$
dependent diamagnetic shift in the field position of the resonance is
observed, Fig. \ref{spectra}a,d. The line also broadens, but the broadening
is roughly three times less than the shift. Below the irreversibility line ($T_{irr}$=31 
K at 1.28 T) the penetration of magnetic flux is hysteretic and
the line position and width are dependent on the direction of magnetic field
sweep. At higher frequencies, above 2.7 T, the CESR line splits in two
components, which are well resolved at 5 K. One of these components is
situated at the position of the normal state CESR. This suggests that
a part of the grains is in the {\it normal state}. In other words, there is a
distribution of $H_{c2}$'s among the grains and in a part of the sample 
$H_{c2}$ is as low as about 2 T.


Figure \ref{diashift} shows the $T$ dependence of the CESR shift, $\Delta
H_{0}(T)=H_{0}(T)-H_{0}(40\,{\rm K})$, at 35 GHz, where $H_{0}(T)$ is the
resonance field of the ESR spectra. It follows closely the field cooled
magnetization $M(T)$ measured at 1.28 T by SQUID magnetometer shown for
comparison.  The rough agreement of $\Delta H_{0}(T)$ and $4\pi $$\left|
M(T)\right| $ and their similar temperature dependences strongly support
that we observe the CESR in the superconducting state. Yet, the CESR line
shift below $T_{c}$ is not simply proportional to the macroscopic
magnetization. Theory and experiment on CESR in superconductors is limited
to a few reports (See Ref.\cite{nemes} and references therein) only.

The complex lineshape seen in Fig. \ref{spectra}e,f at higher fields is
due to an inhomogeneous distribution of $\Delta $H$_{0}$ among the sample
grains. The splitting of the spectra cannot be explained with the variation
of the field between vortex cores. Unlike the NMR spectrum, the CESR is not
broadened by these short-scale magnetic field variations. Electrons diffuse
to large distances within $T_{1}$ and a single resonance appears at a well
defined average field weighted by the local density of states\cite{deGennes1}. 
In K$_{3}$C$_{60}$, a fullerene compound with $T_{c}$=19 K studied in
detail\cite{nemes}, the CESR is relatively narrow in the superconducting
state and only below $T_{irr}$ do large scale (typically 1 $\mu$m)
inhomogeneities of diamagnetism broaden the spectrum.

A substantial part of the spectrum at 2.7 T is not shifted with
respect to the normal state and comes from normal state fractions of the
sample. In other words, $H_{c2}$ of the grains within this fraction is
smaller than the applied field $H$, in this case 2.7 T. The upshifted
line can be identified as signal coming from superconducting parts of the
sample by the similarity of its characteristics to the observed single line
at 1.28 T (Fig. \ref{spectra}d); i.e. $T$ dependent diamagnetic CESR shift and
broadening below $T_{c}$. The value of $\Delta $H$_{0}$ decreases with
increasing field, following the decrease of the superconducting
magnetization and its values at $T$=2.5 K are $\Delta $H$_{0}$(2.7 T)= 25 G, 
$\Delta $H$_{0}$(5.4 T)= 9 G, and $\Delta $H$_{0}$(8.1 T)= 5 G. Thus,
the upshifted line (Fig. \ref{spectra}) arises from particles with large $H_{c2}$. The
diamagnetic shift is larger for particles with larger $H_{c2}$ but since
this has a maximum (at 16 T), the derivative CESR spectrum has a relatively
narrow peak at the high field end. This peak is marked by an arrow in Fig.
\ref{spectra}. As the applied field is increased, the CESR intensity of the
superconducting fraction  with respect to the intensity of the normal state
fraction decreases, Fig. \ref{spectra}e,f. However, a quantitative determination of the variation
of the superconducting fraction is not possible from the CESR lineshape
since the variations of the microwave penetration depth and the spin
susceptibility with magnetic field and temperature are not known.
Nevertheless, our observations provide clear evidence for a low value ($\sim 
$ 2 T) of minimal $H_{c2}$ in the MgB$_{2}$ powder. Since the maximum
measured $H_{c2}$ is about 16 T \cite{wirecondmat}, we conclude that the
sample grains have $H_{c2}$'s spanning from approximately 2 to 16 T.


The anisotropy of MgB$_{2}$ is a probable cause for the distribution of $H_{c2}$'s 
in powder samples. It is unlikely that the distribution is due to a
spread in the quality of our sample, since the superconducting
transition is reproducibly sharp in transport and thermodynamic measurements
and the residual resistance ratio of polycrystalline samples, $RRR>20$, is
relatively high\cite{finnemorePRL}. Still, an
unexpectedly large anisotropy calls for an independent verification. We did this by analysing the
data on the magnetization, $M(H,T)$, of powder samples.

We consider a sample of randomly oriented grains of a uniaxial
superconductor with the anisotropy $\gamma=H_{c2}^{ab}/H_{c2}^{c}$ placed in a field ${\bf H}$
along $z$. The distribution of grains over their $c$ direction is given by 
$dN=N\,\sin \theta \,d\theta /2$ with $\theta $ being the angle between $c$
and ${\bf H}$. The grain upper critical field depends on $\theta $ according
to $H_{c2}(\theta )=H_{c2}^{ab}/\sqrt{\epsilon (\theta )}$ with $\epsilon
=1+(\gamma ^{2}-1)\cos ^{2}\theta $.

In agreement with what is currently known \cite{brasil}, we assume 
$H_{c2}^{ab}>H_{c2}^{c}$ ($\gamma >1$) and consider the field domain
$H_{c2}^{c}<H<H_{c2}^{ab}$ following the procedure of 
\onlinecite{Kyoto}. Clearly, only the grains with $H_{c2}(\theta )>H$
contribute to the superconducting magnetization. The grain orientation 
$\theta _{0}$, for which the given $H$ is the upper critical field is given
by 
$\cos^2\theta_0 = [(H_{c2}^{ab}/H)^2-1]/(\gamma^2-1)$.
We then have for the magnetization 
$M_{z}=\int_{\theta _{0}}^{\pi /2}M_{z}(\theta ,H)\sin \theta \,d\theta $,
while the transverse component of ${\bf M}$ averages to zero. 
According to Ref. \cite{koganPRB}, the magnetization of the grain $\theta $
near its $H_{c2}$ is given by 
\begin{equation}
-4\pi M_{z}=\frac{H_{c2}(\theta )-H}{2\kappa ^{2}\beta \gamma ^{2/3}}%
\,\epsilon (\theta )\,.  \label{mu2}
\end{equation}
Here, $\beta =1.16$ and we assumed the Ginzburg-Landau parameter $\kappa \gg
1$. We then obtain after simple algebra: 
\begin{eqnarray}
M_{z} &=&-M_{0}\,f(h)\,,\quad M_{0}=\frac{\phi _{0}}{32\pi ^{2}\lambda
^{2}\beta \gamma ^{1/3}\sqrt{\gamma ^{2}-1}}\,,  \label{M0} \\
f(h) &=&\frac{1-4h^{2}}{3h^{2}}\,\sqrt{1-h^{2}}+\ln \frac{1+\sqrt{1-h^{2}}}{h%
}\,,  \label{f}
\end{eqnarray}
where $h=H/H_{c2}^{ab}$ and $\lambda =(\lambda _{ab}^{2}\lambda _{c})^{1/3}$ is
the {\it average} penetration depth. It is seen that as $H\rightarrow H_{c2}^{ab}$%
, $M_{z}\propto (H_{c2}^{ab}-H)^{3/2}$, i.e. in a polycrystal $M(H)$ decreases
faster than for a single crystal.

Figure \ref{f1} shows the reversible part of $M(H)$ for a few temperatures,
along with solid curves obtained by fitting the data to Eqs. (\ref{M0}) and (%
\ref{f}). The prefactor $M_0(T)$ and the in-plane upper critical field $%
H_{c2}^{ab}(T)$ are taken as fitting parameters, the best values of which are
shown in the lower panel. By and large, the parameters behave as expected
for $M_0\propto 1/\lambda^2(T)$ and $H_{c2}(T)$, although the low-$T$ value
of $13\,$T for the maximum upper critical field is lower than $\approx 16\,$%
T obtained from the resistivity data \cite{wirecondmat}. The limit $%
M_0(T\rightarrow 0)\approx 0.26\,$G gives $\lambda^2(0)\gamma^{1/3}\sqrt{%
\gamma^2-1}\approx 2.1\times 10^{-9}$cm$^2$. Estimates of $\lambda(0)$ range
between 110 nm\cite{cristen} and 140 nm\cite{finnemorePRL}, which yield $\gamma\approx$ 6--9.


In addition, our analysis of the field dependent resistivity of
MgB$_{2}$ (following the procedure of Ref.\cite{welch} for polycrystalline
superconductors) yields $\gamma \approx$6--9 \cite{budkounpub} in agreement with
values extracted from CESR and M(H,T).

The agreement notwithstanding, one should exercise caution about the large
anisotropy we extract from the magnetization data taken on powder samples.
Our analysis of $M(H,T)$ disregards fluctuations of vortices, the reason
being that MgB$_{2}$ does not seem to have a pronounced structure of weakly
coupled superconducting layers, a prerequisite for strong fluctuations.
Also, we take $M(H)\propto (H_{c2}-H)$ in the whole domain $%
H_{c2}^{c}<H<H_{c2}^{ab}$, too strong an assumption for anisotropies as
large as $\gamma \sim 8$.

In conclusion, CESR shows a large distribution of $H_{c2}$ in high quality MgB$_{2}$ powders.
These results, together with a detailed analysis of magnetization data are suggestive of a significant 
anisotropy of $H_{c2}$. If this is the case, magnetic field dependent experimental
results on MgB$_{2}$ have to be reconsidered. A possible low minimum $H_{c2}\sim 2$ T 
has important consequences on the technical applications of this material. Nevertheless, single
crystals are needed to definitely resolve the issue of anisotropy in MgB$_{2}$.

Support from the Hungarian State Grants, OTKA T029150, and FKFP 0352/1997
and the Swiss National Science Foundation are acknowledged. Ames Laboratory
is operated for the U.S. Department of Energy by Iowa State University under
Contract No. W-7405-Eng-82.


\begin{figure}
\caption{
CESR linewidth versus temperature in MgB$_{2}$ powder at 9 GHz.
The continuous curve is the resistance measured on a pressed pellet from
the same batch as used for CESR. Inset: 9 GHz CESR spectrum at 500 K.
\label{linewidth}}
\end{figure}

\begin{figure}
\caption{
CESR spectra of MgB$_{2}$ at various frequencies a) - c) in the
normal state at 40 K; and d) -f) at $T$=5 K. The whole sample is
superconducting at $H$=1.28 T and 5 K. Note the diamagnetic shift of the
resonance with respect to the 40 K spectrum. Part of the sample is in the
normal state at higher ESR frequencies, the superconducting component is
marked by an arrow. Dashed and full lines below the experimental spectra are
Lorentzian fits to the normal and superconducting CESR, respectively.
\label{spectra}}
\end{figure}

\begin{figure}
\caption{
Temperature dependence of the diamagnetic shift (full symbols are 
up, open symbols are down sweeps) of the CESR at 35 GHz (1.28 T) and diamagnetic magnetization 
measured by SQUID (solid curve) at 1.28 T.
\label{diashift}}
\end{figure}

\begin{figure}
\caption{
The upper panel shows the reversible magnetization $M(H)$ of
the MgB$_2$ powder at temperatures from 6 to 34 K with a 2 K step. Solid
lines are calculated with the help of Eqs. (\ref{M0}) and (\ref{f}) with two
fitting parameters $M_0$ and $H_{c2}^{ab}$. The latter are plotted in the lower panel.
\label{f1}}
\end{figure}


\begin{references}
\bibitem[*]{lapertot}  On leave from Commissariat \`{a} l'Energie Atomique,
DRFMC-SPSMS, 38054 Grenoble, France.

\bibitem{akimitsu}  J. Nagamatsu {\it et al.}, Nature, {\bf 410}, 63 (2001).


\bibitem{finnemorePRL}  D. K. Finnemore {\it et al.}, Phys. Rev. Lett., {\bf %
86}, 2420 (2001).

\bibitem{wirecondmat}  S. L. Bud'ko {\it et al.}, Cond-mat/0102413.

\bibitem{junod}  Y. Wang, T. Plackowski, A. Junod, Cond-mat/0103181.

\bibitem{hinksSH}  F. Bouquet {\it et al.}, Cond-mat/0104206.

\bibitem{shulga}  S. V. Shulga {\it et al.}, cond-mat/0103154.

\bibitem{bascones}  E. Bascones, F. Guinea, cond-mat/0103190.

\bibitem{liu}  A. Y. Liu, I. I. Mazin, and J. Kortus, cond-mat/0103570.

\bibitem{haas}  S. Haas, K. Maki, cond-mat/0104207.

\bibitem{chen}  C - T. Chen {\it et al.}, cond-mat/0104281.

\bibitem{pronin}  A. V. Pronin {\it et al.}, cond-mat/0104291.

\bibitem{brasil}  O. F. de Lima {\it et al.}, cond-mat/0103287.

\bibitem{patnaik}  S. Patnaik {\it et al.}, unpublished.

\bibitem{budkoPRL}  S. L. Bud'ko {\it et al.}, Phys. Rev. Lett. {\bf 86},
1877 (2001).

\bibitem{andrasphysC}  A. Janossy, R. Chicault, Physica C(Amsterdam), {\bf %
192}, 399 (1992).

\bibitem{mazin}  J. Kortus {\it et al.}, Cond-mat/0101446.

\bibitem{canfieldPRL}  P. C. Canfield {\it et al.}, Phys. Rev. Lett., {\bf 86}%
, 2423 (2001).

\bibitem{winter}  J. Winter: Magnetic Resonance in Metals, Clarendon Press,
(Oxford), 1970.

\bibitem{yafet}  Y. Yafet, Solid State Phys., {\bf 14}, 1 (1963).

\bibitem{nemes}  N. M. Nemes {\it et al.}, Phys. Rev. B, {\bf 61}, 7118
(2000).

\bibitem{deGennes1}  P. G. de Gennes, Solid State Comm., {\bf 4}, 95 (1966).

\bibitem{Kyoto}  V.G. Kogan and J.R. Clem, Proc. of LT-18, Jap. Journ. Appl.
Physics, {\bf 26}, 1159 (1987).

\bibitem{koganPRB}  V.G. Kogan and J.R. Clem, Phys. Rev. B, {\bf 24}, 2497
(1981).

\bibitem{cristen}  J. R. Thompson {\it et al.}, cond-mat/0103514.

\bibitem{welch}  D. O. Welch et al., Phys. Rev. B, {\bf 36}, 2390 (1987).

\bibitem{budkounpub}  S. L. Bud'ko {\it et al.}, unpublished.
\end{references}
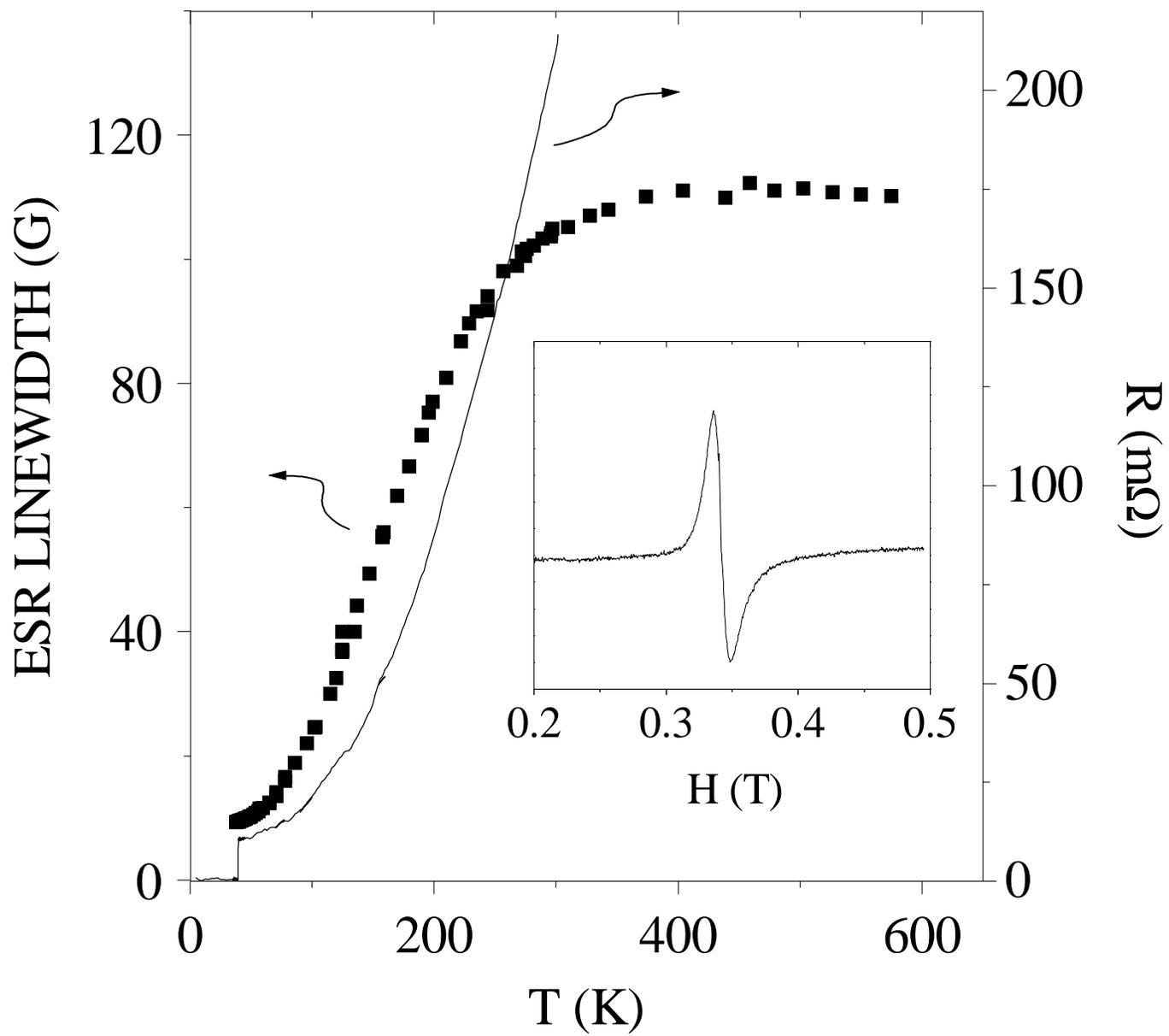
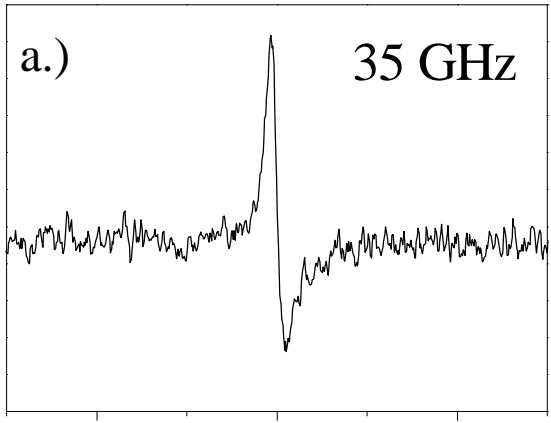
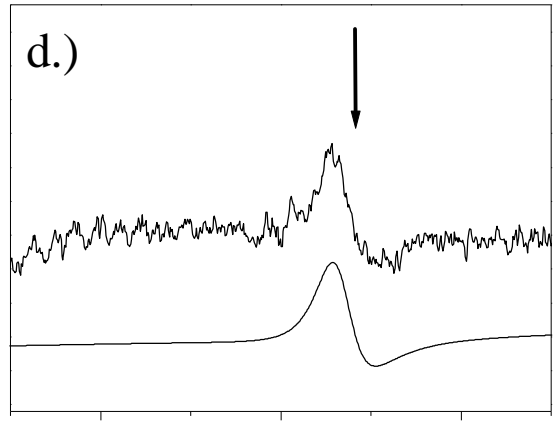
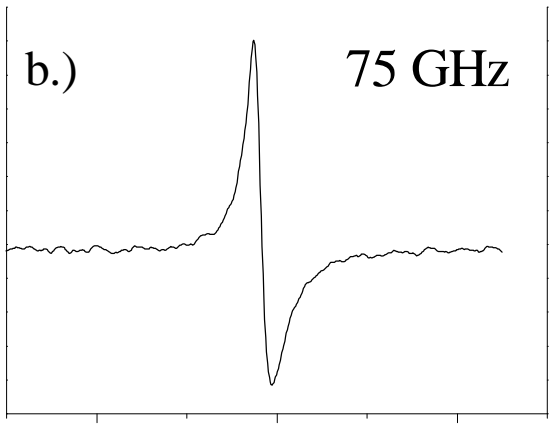
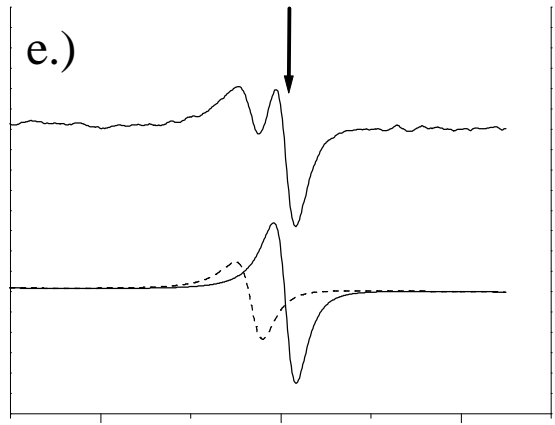
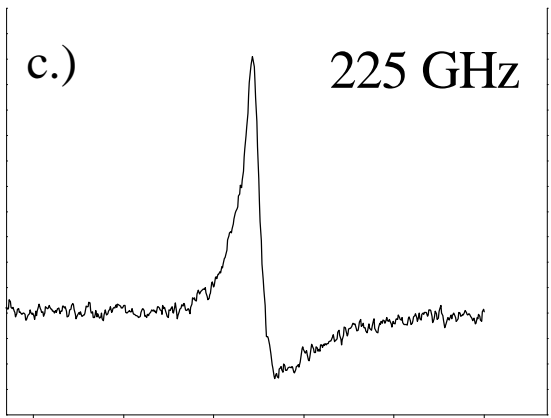
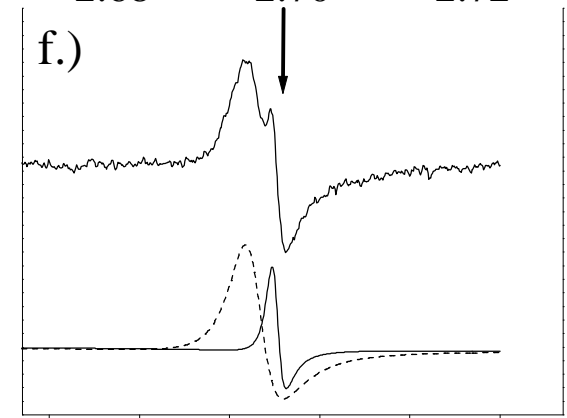
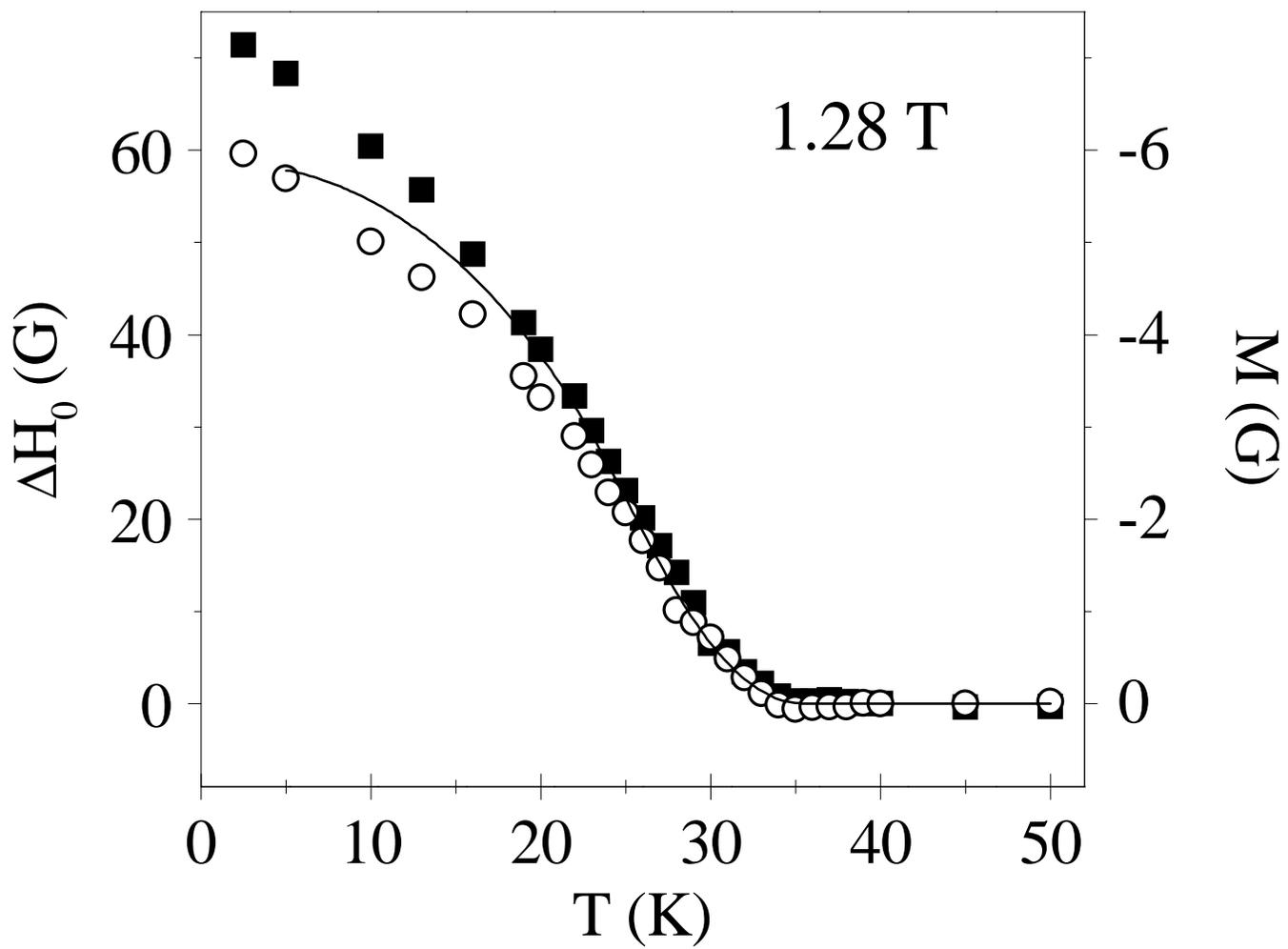
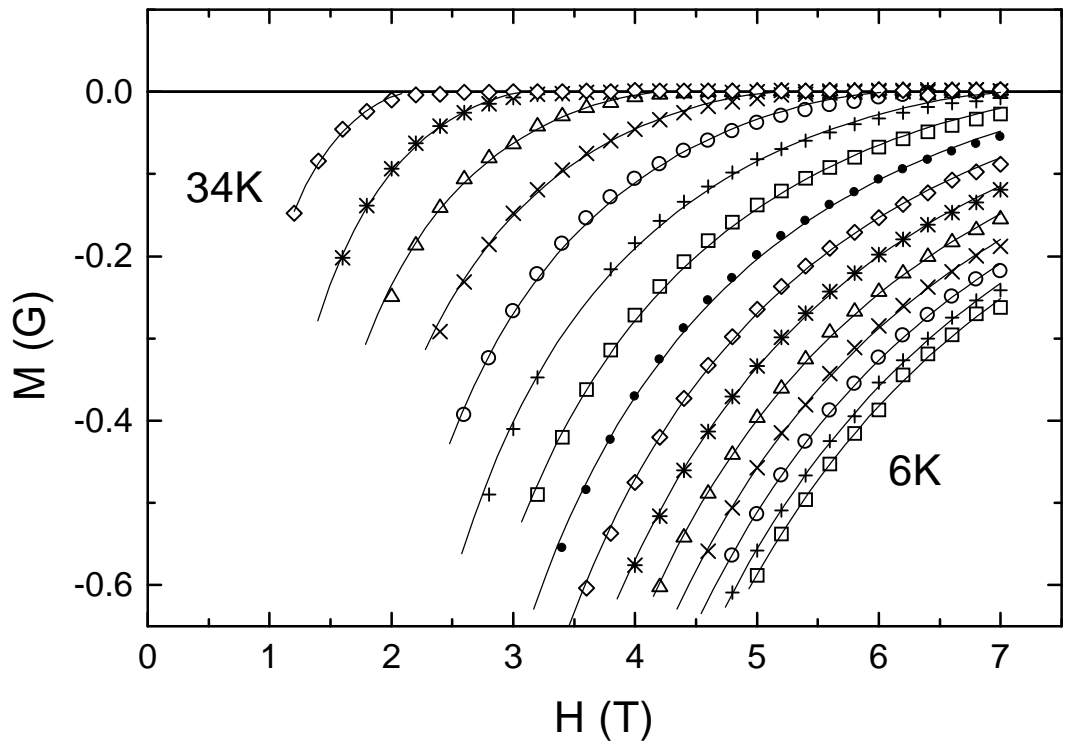
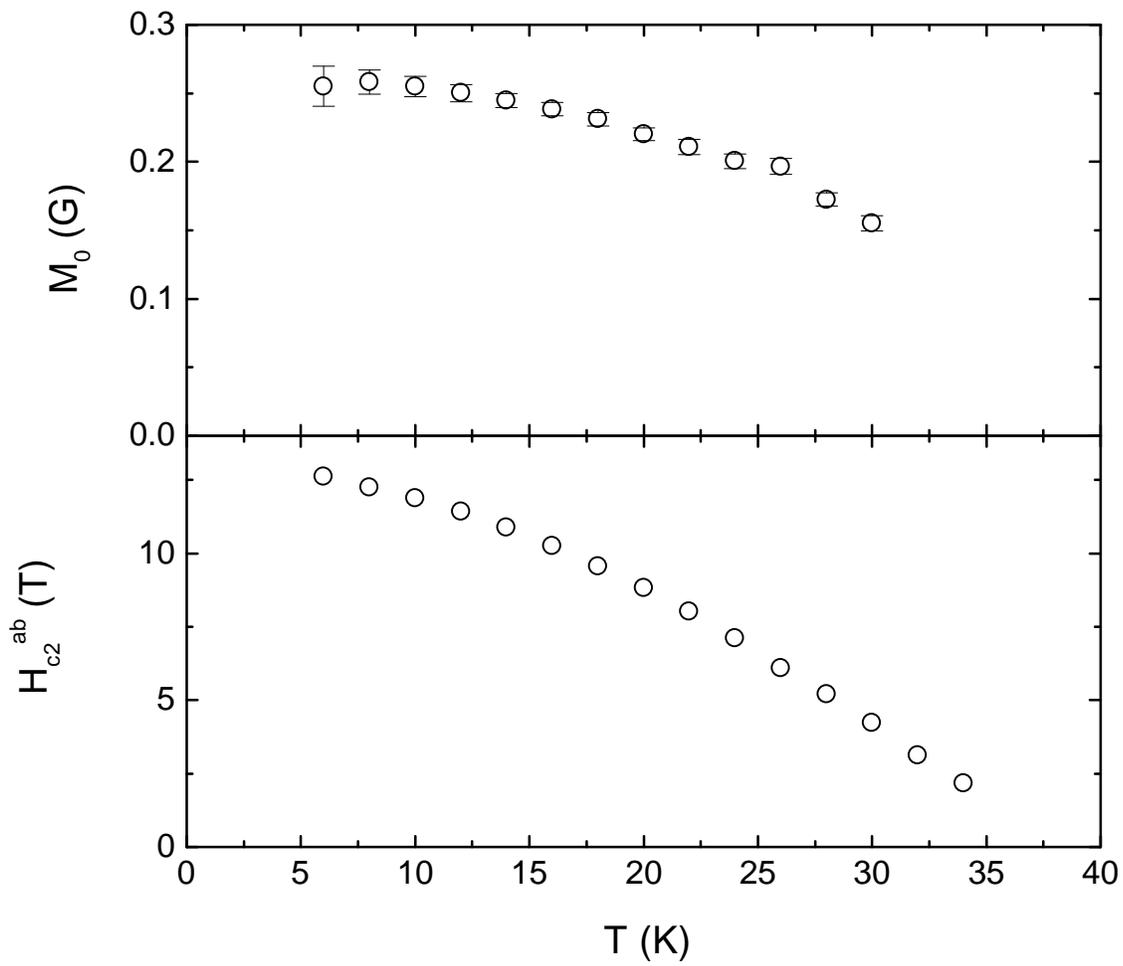
\end{document}